\documentclass[showkeys,
reprint,
superscriptaddress,
 amsmath,amssymb,
 aps,
]{revtex4-2}

\usepackage{graphicx}
\usepackage{amsfonts}
\usepackage{xcolor}

\DeclareMathOperator{\re}{\mathop{\mathrm{Re}}}

\begin{document}

\preprint{APS/123-QED}

\title{Miniaturization of Josephson junction for digital superconducting circuits}

\author{I.~I.~Soloviev}
\email{isol@phys.msu.ru}
\affiliation{Lomonosov Moscow State
	University Skobeltsyn Institute of Nuclear Physics, 119991 Moscow,
	Russia} 
\affiliation{Dukhov All-Russia Research Institute of Automatics, Moscow 101000, Russia}
\affiliation{Physics Department of Moscow State University, 119991 Moscow, Russia}

\author{S.~V.~Bakurskiy}
\affiliation{Lomonosov Moscow State
	University Skobeltsyn Institute of Nuclear Physics, 119991 Moscow,
	Russia} 
\affiliation{Dukhov All-Russia Research Institute of Automatics, Moscow 101000, Russia}
\affiliation{Moscow Institute of Physics and Technology, State
	University, 141700 Dolgoprudniy, Moscow region, Russia}

\author{V.~I.~Ruzhickiy}
\affiliation{Lomonosov Moscow State
	University Skobeltsyn Institute of Nuclear Physics, 119991 Moscow,
	Russia}
\affiliation{Dukhov All-Russia Research Institute of Automatics, Moscow 101000, Russia}
\affiliation{Physics Department of Moscow State
	University, 119991 Moscow, Russia}

\author{N.~V.~Klenov}
\affiliation{Lomonosov Moscow State
	University Skobeltsyn Institute of Nuclear Physics, 119991 Moscow,
	Russia} 
\affiliation{Dukhov All-Russia Research Institute of Automatics, Moscow 101000, Russia}
\affiliation{Physics Department of Moscow State
	University, 119991 Moscow,
	Russia}

\author{M.~Yu.~Kupriyanov}
\affiliation{Lomonosov Moscow State
	University Skobeltsyn Institute of Nuclear Physics, 119991 Moscow,
	Russia} 

\author{A.~A.~Golubov}
\email{a.a.golubov@utwente.nl}
\affiliation{Moscow Institute of Physics and Technology, State
	University, 141700 Dolgoprudniy, Moscow region, Russia}
\affiliation{Faculty of Science and Technology and MESA+ Institute of Nanotechnology, 7500 AE Enschede, The Netherlands}

\author{O.~V.~Skryabina}
\affiliation{Lomonosov Moscow State
	University Skobeltsyn Institute of Nuclear Physics, 119991 Moscow,
	Russia}
\affiliation{Moscow Institute of Physics and Technology, State
	University, 141700 Dolgoprudniy, Moscow region, Russia}

\author{V.~S.~Stolyarov}
\affiliation{Moscow Institute of Physics and Technology, State
	University, 141700 Dolgoprudniy, Moscow region, Russia}
\affiliation{Dukhov All-Russia Research Institute of Automatics, Moscow 101000, Russia}

\date{\today}

\begin{abstract}
In this work we briefly overview various options for Josephson junctions which should be scalable down to nanometer range for utilization in nanoscale digital superconducting technology. Such junctions should possess high values of critical current, $I_c$, and normal state resistance, $R_n$. Another requirement is the high reproducibility of the junction parameters across a wafer in a fabrication process. We argue that Superconductor - Normal metal - Superconductor (SN-N-NS) Josephson junction of ``variable thickness bridge'' geometry is a promising choice to meet these requirements. Theoretical analysis of SN-N-NS junction is performed in the case where the distance between the S-electrodes is comparable to the coherence length of the N-material. The restriction on the junction geometrical parameters providing the existence of superconductivity in the S-electrodes is derived for the current flowing through the junction of an order of $I_c$. The junction heating as well as available mechanisms for the heat removal is analyzed. The obtained results show that an SN-N-NS junction with a high (sub-millivolt) value of  $I_cR_n$ product can be fabricated from a broadly utilized combination of materials like Nb/Cu using well-established technological processes. The junction area can be scaled down to that of semiconductor transistors fabricated in the frame of a 40-nm process.

\end{abstract}

\maketitle

\section{Introduction}

The promised end of Moore's law \cite{Moore} brings attention to the so-called ``beyond complementary-metal-oxide semiconductor (CMOS)'' technologies. One of them is the superconductor technology distinguished by high energy efficiency and high clock frequencies \cite{Holmes,Tolp,Beil,IRDS2020}. It is considered to be especially suitable for ``cold electronics'' operating in the gradient between room temperature and temperature of cryogenic payloads like quantum computers, quantum internet, or scalable sensors \cite{DeBen}. Since the 1980s, superconducting circuits have been consistently developed reaching a fairly mature level at the end of the 1990s, showing an implementation of digital and mixed-signal devices \cite{Mukh1,Mukh2,Mukh3}.

However, further progress has slowed down. The main reason was the low integration density. It caused, e.g., implementations of superconducting memory to be of impractical storage capacity \cite{SQMem} that in turn impeded the realization of superconducting processors. Nevertheless, a decade later the researches were whipped up by US government investments aimed at the development of a notional prototype of a superconducting computer \cite{IARPA}. Its implementation is to show a prospective application of superconductor technology in the field of supercomputing. Unfortunately, along with significant overall progress in design and fabrication, we have to admit that the functional density of superconducting circuits is still quite low. The recently demonstrated benchmark circuits for the modern state-of the-art 150~nm of Massachusetts Institute of Technology Lincoln Laboratory (MIT LL) processes are the shift registers with $1.3 \times 10^7$ Josephson junctions (JJs) per square centimeter circuit density \cite{TolpygoASC}. This can be compared with the achieved $\sim 1.3 \times 10^{10}$~Tr/cm$^2$ density of transistors in the most advanced 5~nm process node of the metal–oxide–semiconductor field-effect transistor (MOSFET) technology.

One of the most difficult puzzles of the scaling is the reduction of the size of the Josephson junction \cite{Tolp}, which is a nonlinear element of superconducting circuits. In comparison with the modern transistor, the area of the junction is more than two orders of magnitude greater nowadays.

In this paper, we examine the most common types of Josephson junctions on their scalability. We argue that a planar SN-N-NS junction (S stays for a superconductor, and N - for a normal metal) with variable-thickness bridge geometry is promising for miniaturization. We calculate temperature dependencies of the supercurrent and the characteristic voltage of the SN-N-NS junction for various transparencies of the SN interfaces. Further, we consider limitations on the junction critical current providing stable  superconducting state in the S-electrodes. We also discuss the heat balance in the SN-N-NS structure and estimate the heating during junction operation. Finally, we discuss the achievable range of areas of the studied structures and possible ways of their fabrication.

\section{Scalability of Josephson junctions}

A workhorse of digital superconductor technology is a sandwich Superconductor/Insulator/Superconductor (SIS) Nb/Al-AlO$_x$/Nb Josephson junction. Due to the good wetting of Nb with Al, one can obtain a high uniformity of barrier transparency which provides just a few percents of the technological spread \cite{MITech} of its parameters: critical current, $I_c$, and normal state resistance, $R_n$. The routinely achieved critical current density of the junctions, $j_c = 0.1$~mA/$\mu$m$^2$, corresponds to a small insulator thickness still providing the homogeneous SI boundary. The critical current of Josephson junction should exceed the noise current, $I_n = (2\pi/\Phi_0) k_B T$ (where $k_B$ is the Boltzmann constant, $T$ is the temperature, $\Phi_0 = \pi\hbar/e$ is the flux quantum, $\hbar$ is the reduced Planck constant and $e$ is the electron charge), by about three orders of magnitude for low enough bit error rate. At the same time, the energy dissipation during junction switching is proportional to the critical current, $E_J \approx I_c\Phi_0$, that makes its high values undesirable. For standardly used helium temperature, $T = 4.2$~K, it turns to $I_c \approx 0.1$~mA, and area of the junction, $a = I_c / j_c \approx 1~\mu$m$^2$.

Since the sandwich type of tunnel Josephson junction possesses relatively high self-capacitance, $c \geq 60$~fF/$\mu$m$^2$, the resistive shunt, $R_s$, is always used to damp Josephson oscillations occurring after the junction switching. A workable operation regime is achieved with Steward-McCumber parameter value, $\beta_c = (2\pi/\Phi_0) j_c c R^2_s a^2 \approx 1$. Corresponding $R_s \approx 5$~$\Omega$ is commonly implemented with MoN$_x$ having resistance $\sim 5$~$\Omega$ per square \cite{MoNx}. Taking into account interconnect area with minimum wiring feature size (ca. 0.5-1~$\mu$m) the shunt about triples the total area of the junction. The parallel combination of the shunt and the internal resistance provides sum-mV characteristic voltage and sub-THz characteristic frequency of the junction .

A natural approach to scale the junction area below micron size is to increase the critical current density. Starting from $j_c = 0.5$~mA/$\mu$m$^2 $ the junction becomes self-shunted \cite{TolpSSJ,TolpSSJ1}. For example, the most functionally dense Random-Access-Memory (RAM) circuit fabricated recently \cite{STMem} were based on the Josephson junctions having $j_c = 0.6$~mA/$\mu$m$^2$ with no shunt resistors \cite{TolpSSJ1}, whilst minimum junction area was about $a \approx 0.2~\mu$m$^2$ ($\beta_c \approx 2$). However, the critical current density increase corresponds to the decrease of tunneling barrier thickness and according increase of its inhomogeneity in coordinate and momentum space. The influence of fluctuations in AlO$_x$ transparency (caused, e.g., by pin holes and localized states) on spread of Josephson junction critical currents becomes even more pronounced with the decrease of the junctions area . Since complex circuits require the uniformity of junctions, this scaling strategy is practically limited to the junction size of a few tenths of a micron.

Double-barrier, SINIS, Josephson structures were proposed to mitigate the difficulties of the scaling \cite{DBJJ,BCGKSR2001}. If the N-layer thickness, $d_n$, is much smaller then the decay length, 
\begin{equation}
d_n\ll \xi_{n}=\sqrt{\hbar D_{n}/2 \pi k_B T_{c}}
\end{equation}
(dirty limit), where $T_{c}$ is the superconductor critical temperature, $D_{n}=v_{F}l/3$ is the diffusion coefficient, $v_{F}$ is the Fermi velocity and $l$ is the electron mean free path in the normal metal, then the cotunneling across INI weak link may prevail over sequential SIN$+$NIS tunneling process resulting in $I_{c}R_{n}$ product even larger than that of SIS junction with the same $I$ barrier \cite{kuprlukichevA,galaktionov_zaikin2002,SINISnote}. Simple estimations show \cite{DBJJ} that even the formation of small pin holes of a diameter $d_{ph}\ll \xi_{n}, ~d_n$, will not lead to large spread of the junction parameters if $\sqrt{d_{ph}\xi_{n}}\ll d_n$. 

Unfortunately, due to difference in Al growth over Nb and AlO$_{x}$ surfaces, the implementation of equal barriers in Nb/AlO$_{x}$/Al/AlO$_{x}$/Nb structure appeared to be intractable task with increase of the barriers transparency \cite{BCIKSBGKR2003,TBGK2003}. The structure asymmetry led to localization of the weak link at one of the barriers that cancelled all the benefits, making this approach impractical for scaling \cite{TBGK2003}. The implementation of the much-needed symmetry of the transparency of the tunnel layers required the development of two different technological processes of aluminum oxidation. In the absence of such technology, the SINIS structures turned out to be useful only for implementation of Josephson voltage standards \cite{PhysC2002,Behr2005,Mueller2007}.

Another option is to artificially synthesize the interlayer material by doping semiconductor up to a degenerate state \cite{SDS,GKL1988,Kulikov1991,Baek2006,Baek2007,Olaya2008,Olaya2009,Chong2009,Muller2013,Olaya2014,Olaya2015,Cao2018,Olaya2019,Haygood2019,Olaya2019a}, e.g., using $\alpha$-Si, where Nb or W can be chosen as dopants. 
The current transport in Nb/$\alpha$-Si/Nb (SDS, where D stays for doped semiconductor) junction is determined by elastic and inelastic resonance tunneling processes. The latter form quasi-one-dimensional channels with metallic conductivity and provide internal junction shunting. While $I_{c}R_{n}$ product of sub-millivolt level is readily attainable \cite{SDS,Olaya2015,Olaya2019,Haygood2019,Olaya2019a}, the inherent inhomogeneity of the barrier transparency prohibits the scaling of such SDS structure with preservation of the junction parameters reproducibility due to the probabilistic nature of the formation of resonant channels for the flow of normal and superconducting currents.

At nanometer scale, where area of Josephson junction is $a \sim 0.01~\mu$m$^2$, the Josephson junction critical current should be $j_c \sim 10$~mA/$\mu$m$^2$, taking into account the required critical current $I_c = 0.1$~mA at the working temperature $T = 4.2$~K. Such high values of the critical current density together with the considered inhomogeneity of the barriers of SIS and SDS junctions makes SNS junction to be the preferred type for scaling.

SNS structure obviously does not require shunt resistor. It is important to choose N-material so as to minimize the suppression of superconductivity in S-electrodes while maximizing the induction of superconductivity in N-interlayer for high values of $I_{c}R_{n}$ product and critical current, correspondingly. For transparent SN interface the proximity effect is described \cite{Ivanov1981,Kupriyanov1982} by the suppression parameter, 
\begin{equation}
\gamma =\frac{\rho_{s}\xi_{s}}{\rho_{n}\xi_{n}},~   \xi_{s}=\sqrt{\frac{\hbar D_{s}}{2 \pi k_B T_{c}}},
\end{equation}
where $\xi_s$ is decay length in S-material and $\rho_{s,n}$ are normal resistivities of S- and N-materials. The desired value, $\gamma <1$, means that the amount of normal electrons diffusing per second from N to S is smaller than the same value of correlated electrons moving in the opposite direction. This case is commonly realized with pairs of materials like Nb (S) and Ti, Hf or Pd$_{x}$Au$_{1-x}$ (N), where $\xi_{n}\lesssim \xi _{s}\approx 10$~nm while resistivity of normal metals is larger or of an order of the one of niobium, $\rho _{n}\gtrsim \rho_{s}\approx 8~\mu \Omega ~$cm. For sandwich type  SNS junction and the N layer thickness larger than $\xi_{n}$, it is necessary \cite{Kupriyanov1982} to have $\gamma \lesssim 0.1$ for achieving a high value of $I_{c}R_{n}$ product at $T/T_{c}\approx 0.5$. However, this value of suppression parameter is difficult to implement. It is, for this reason, the experimentally obtained magnitudes of characteristic voltage is rather small \cite{DBJJ,SNS,PhysC2000SNS,Hagedorn2001,Schubert2001,Hagedorn2002,Dresselhaus2003,Hagedorn2003,schubertt2005,Hagedorn2006,Nagel2011}. Note that a simple decrease in the N layer thickness is not a solution of this problem. In addition to technological difficulties, the decrease in N layer thickness leads to spatial delocalization of the weak link region due to the depairing process in the vicinity of the interfaces \cite{Ivanov1981,Zubkov1983,Kupriyanov1992}.

\begin{figure}[t]
\includegraphics[width=0.8\columnwidth,keepaspectratio]{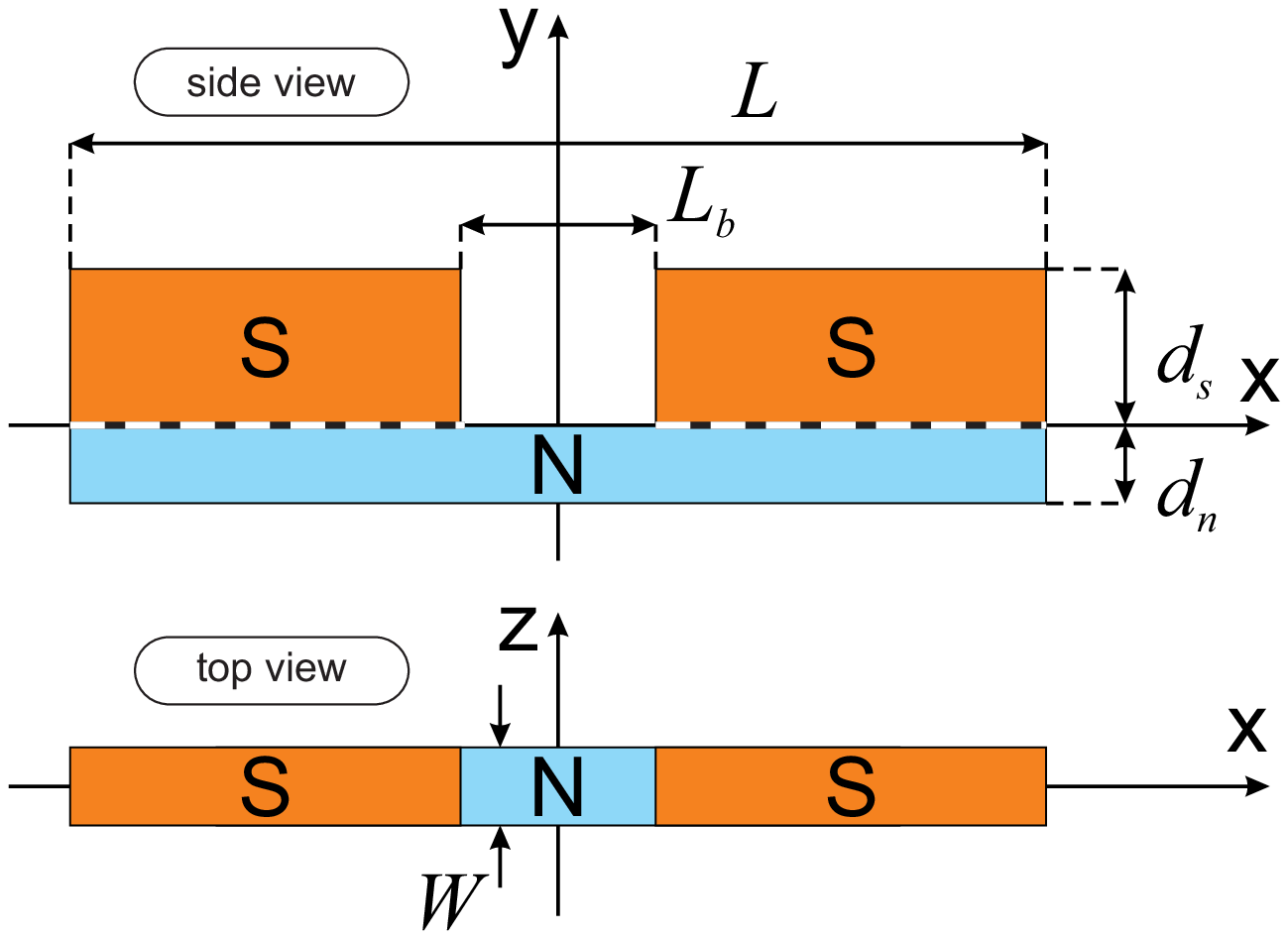}  
\caption{Sketch of SN-N-NS Josephson junction with variable thickness bridge geometry.}
\label{Fig1}
\end{figure}

This problem is inherent for commonly used sandwich-type Josephson junctions. However, it can be circumvented by modification of the weak link region geometry with corresponding redistribution of current flowing therein \cite{Likharev1971,Likharev1975,Likharev1976,Zaikin2001,Vodolazov2020}. An example of such solution is the SN-N-NS junction in the form of variable thickness bridge \cite{Warlaumont1979,Beasley1981,Liengme1983,Nakano1987,Lukens1985,Baryshev1989a,Baryshev1989} presented in Fig.~\ref{Fig1}. Here a high current density in the bridge is lowered due to the current spread all over the area under the electrodes. The latter is owing to the finite transparency of SN interfaces \cite{Lukens1985,Zehnder1999,Pratt1999,Pratt2000}.

Metals with relatively large decay length are preferred to obtain high critical current value of SN-N-NS junction. Unfortunately, such metals generally possess relatively small resistivity, since $\xi_{n}\sim \sqrt{D_{n}}$ while $\rho_{n}\sim 1/D_{n}$. However, if the thickness of N layer is smaller than the decay length, $d_{n}\ll \xi_{n}$, then the suppression parameter is decreased \cite{Golubov1983} proportionally to their ratio,
\begin{equation}
\gamma_{M}=\gamma \frac{d_{n}}{\xi_{n}}, 
\end{equation}
corresponding to the decreased number of normal electrons diffusing from N to S. Thus, the SN-N-NS structure size ratio, $d_{n}\ll \xi_{n} \approx L_{b}$, seems to be optimal choice to provide both: high critical current and normal state resistance of the junction.

Below we consider SN-N-NS junction with specified ratio of geometrical parameters as the promising candidate for scaling, since the decay length, $\xi_n$, of the broadly utilized metals, like Al or Cu, lies in the range of a few tens of nanometers. 

\section{Calculation of SN-N-NS junction supercurrent}

Our model of SN-N-NS junction contains a normal metal film connecting two massive superconducting electrodes of the length, $(L-L_b)/2$ each, located at the distance $\pm L_b/2$ from the center of this film, see Fig.~\ref{Fig1}. The total length of the junction is $L$. In the calculation of the critical current, we suppose that condition of dirty limit is fulfilled for all metals, the critical temperature of N material is equal to zero and its width, $W$, and thickness, $d_n$, are much smaller than Josephson penetration depth, $\lambda_J$ , and decay length, $\xi_n$, respectively.

The proximity effect in this system can be considered in the frame of Usadel equations \cite{Usadel}, which in the N film have the form,
\begin{equation}
\xi^{2}_n \frac{\partial }{\partial x}\left( G_{n}^{2}\frac{\partial \Phi_{n}}{\partial x}\right) +\xi^{2}_n\frac{\partial }{\partial y}\left( G_{n}^{2}\frac{\partial \Phi_{n}}{\partial y}\right) =\omega G_{n}\Phi_{n}.
\label{equTetaNF}
\end{equation}
Here $\Phi_{n},$ $G_{n}=\omega /\sqrt{\omega ^{2}+\Phi_n\Phi_{n}^{\ast }}$ are modified Usadel Green's functions, and $\omega =(2m+1)T/T_{c}$ are Matsubara frequencies, $m$ is integer. $\Phi_{n},  $ are normalized on $\pi T_c$.

We apply standard boundary conditions at the edges of the structure \cite{Suppl} and Kupriyanov-Lukichev boundary condition \cite{kuprlukichevA} at SN interfaces. The conditions, $\gamma_{M}\lesssim 0.3$, $d_{n}\ll \xi_{n}$ \cite{Golubov1983}, permit us to neglect the suppression of superconductivity in the S film due to proximity effect even in the case of fully transparent SN interfaces, and also allows considering functions $\Phi_{n}$ independent on the coordinate $y$, in the first approximation on the parameter $d_{n}/\xi_n$. 

By integration of equations (\ref{equTetaNF}) over the coordinate $y$, we obtain the equations,
\begin{equation}
\frac{\gamma_{BM}\xi^{2}_n }{G_{n}}\frac{\partial }{\partial x}\left( G_{n}^{2}\frac{\partial \Phi_{n}}{\partial x}\right) -\left( G_{s}+\gamma_{BM}\omega\right) \Phi_{n}=-G_{s}\Phi_{s},  \label{EqUnderSr}
\end{equation}
for the region of electrodes, $|L_b/2|\leq |x|\leq |L/2|$, where $\gamma_{BM}=\gamma_{B}d_{n}/\xi_{n} $, $\gamma_{B}=R_{B}/\rho_{n}\xi_{n}$ is suppression parameter, $R_{B}$ is specific resistance of SN interface, $\Phi_{s}, $ are modified Usadel Green's functions normalized on $\pi T_c$, $G_{s}=\omega /\sqrt{\omega ^{2}+\Phi_s\Phi_{s}^{\ast }}$. In the area of the bridge we have the following equations,
\begin{equation}
\xi^{2}_n \frac{\partial }{\partial x}\left( G_{n}^{2}\frac{\partial \Phi_{n}}{\partial x}\right) =\omega G_{n}\Phi_{n}.  \label{EqOutS}
\end{equation}

The solution of the problem is simplified in the limit of a small gap between superconducting electrodes, $L_b\ll \xi_{n} $. We also suppose that in the practically interesting case, the length of superconducting electrodes is much larger than the characteristic scale, $L-L_b \gg \zeta$, which arise in the frame of analytical solution of the problem \cite{Suppl}. This characteristic scale is $\zeta \approx \sqrt{\gamma_{BM}} \xi_{n} $  in the limit of small $\gamma_{BM}$ for arbitrary normalized temperature, $t = T/T_c$. In the opposite limit of large $\gamma_{BM}$, it is  $\zeta \approx \xi_{n}$ at small temperatures and $\zeta \lesssim \xi_{n}$ at $t\gtrsim 0.5$. 

Under these assumptions, we obtain \cite{Suppl} the following expression for the product of superconducting current, $J_s$, across the junction and its normal state resistance, $R_n$,

\begin{equation}
\frac{eR_{n}J_{s}}{2\pi k_B T_{c}}=t\sum_{\omega \geq 0}\frac{2\re\Phi_{n}\left(1+2\eta \right) }{\sqrt{\omega^{2}+(\re\Phi_{n})^{2}}}C,
\label{Js1}
\end{equation}
where $\eta = \sqrt{\gamma_{BM}} \xi_{n}/L_b$, C are the constants determined by matching solutions in the ranges of electrodes and the bridge \cite{Suppl}, and $\re\Phi_n$ is the real part of the functions $\Phi_n$, while
\begin{equation}
R_{n}=R_{nb}+2R_{sn} \label{Rjun}
\end{equation}
is the sum \cite{Karminskaya2010} of the resistance of the bridge, $R_{nb}$, and the resistances of two SN interfaces, $R_{sn}$, 
\begin{equation*}
R_{nb}=\frac{\rho_{n}L_b}{d_{n}W},~R_{sn}=\frac{\rho_{n}\xi_{n}\sqrt{\gamma_{BM}}}{d_{n}W}.
\end{equation*}

In the limit of small suppression parameter, 
\begin{equation*}
\sqrt{\gamma_{BM}} \ll L_b/\xi_{n} \ll 1
\end{equation*}
(rigid boundary conditions), the expression (\ref{Js1}) for supercurrent takes the following form, 
\begin{equation}
\frac{eR_{n}J_{s}}{2\pi k_B T_{c}}=t\sum_{\omega \geq 0}\frac{2\Delta \cos \frac{\varphi }{2}}{\Omega_{1}}\arctan \frac{\Delta \sin \frac{\varphi }{2}}{\Omega_{1}},  \label{Js1KO}
\end{equation}
where $R_{n} = R_{nb}$, $\Delta$ is the magnitude of superconducting order parameter normalized on $\pi T_{c}$, $\varphi$ is superconducting phase difference across the junction, and
\begin{equation*}
\Omega_{1}=\sqrt{\Omega^{2}+\Delta^{2}\cos^{2}\frac{\varphi}{2}},~\Omega = \omega\left(1+\gamma_{BM}\sqrt{\omega^{2}+\Delta^{2}}\right).
\end{equation*}
For a vanishingly small suppression, $\gamma_{BM}\rightarrow 0$, the expression (\ref{Js1KO}) transforms into the formula obtained by Kulik and Omelyanchuk (KO-1) \cite{KO1}.

In the opposite limit of large $\gamma_{BM}$, 
\begin{equation*}
\frac{L_b}{\xi_{n}} \ll \frac{\gamma_{BM}}{(1+\gamma_{BM})},
\end{equation*}
expression (\ref{Js1}) transforms into
\begin{equation}
\frac{eR_{n}J_{s}}{2\pi k_B T_{c}}=t\sum_{\omega \geq 0}\frac{\sqrt{2}\Delta^{2}\sin \varphi}{\Omega_{1}\sqrt{\left(\sqrt{\Omega^{2}+\Delta^{2}}+\Omega_{1}\right)\sqrt{\omega^{2}+\Delta^{2}}}},  \label{current}
\end{equation}
where $R_{n}$ is mainly determined by the resistance of SN interfaces.

\section{$I_cR_n$ product of SN-N-NS junction}

\begin{figure}[b]
	\includegraphics[width=1\columnwidth,keepaspectratio]{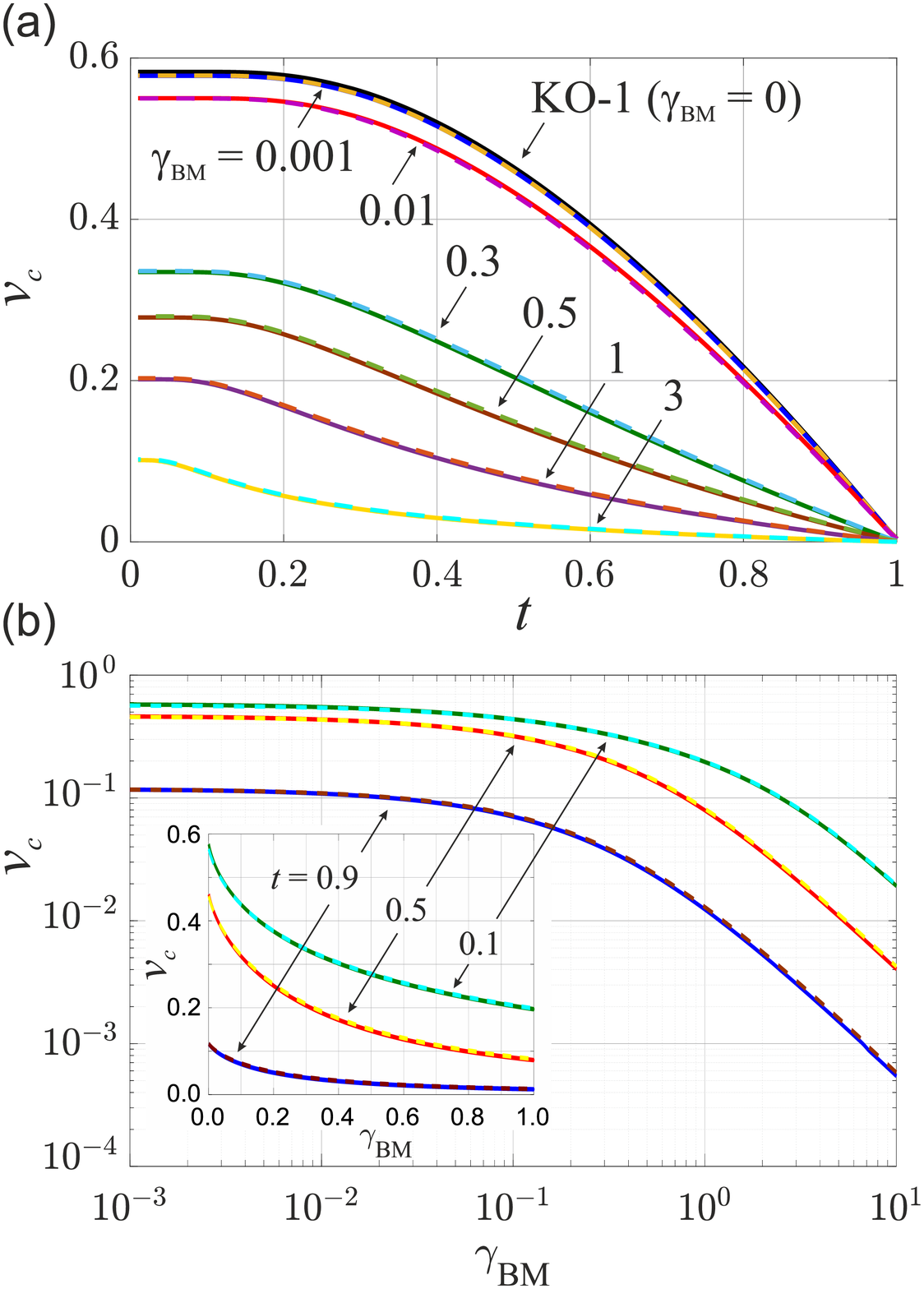}
	\caption{Normalized $I_cR_n$ product, $v_c = eI_cR_n/2\pi k_BT_c$, of SN-N-NS Josephson junction in dependence on (a) normalized temperature, $t$, and (b) 		suppression parameter, $\gamma_{BM}$, calculated using (\protect\ref{Js1}) (solid lines), (\ref{Js1KO}) (dotted lines, $\gamma_{BM} = 0.001,~0.01$, panel (a)) and (\protect\ref{current}) (dotted lines, $\gamma_{BM} = 0.3 ... 3$, panel (a) and panel (b)). Inset shows $v_c(\gamma_{BM})$ for $\gamma_{BM} \leq 1$. The upper solid line in	panel (a) corresponds to KO-1 expression \cite{KO1}.}
	\label{Fig2}
\end{figure}

Characteristic voltage of SN-N-NS junction can be obtained directly from expression (\ref{Js1}) and its limiting cases (\ref{Js1KO}), (\ref{current}). Fig.~\ref{Fig2}(a) shows normalized $I_cR_n$ product, $v_c = eI_cR_n/2\pi k_B T_c$  (where $I_c = \max[J_s(\varphi)]$), as a function of normalized temperature, $t$. The normalizing coefficient value is $2\pi k_B T_c/e \approx 5$~mV for niobium critical temperature, $T_c = 9.2$~K. The bridge length is $L_b = 0.1 \xi_n$. The curves obtained in both limits are well consistent. The presented dependencies are limited from above by the curve obtained using KO-1 expression \cite{KO1} for $\gamma_{BM} = 0$ as expected. The temperature corresponding to characteristic voltage drop by 20\% from its maximum value decreases from $t \approx 0.5$ at small suppression, $\gamma_{BM} \le 0.01$, to $t \le 0.2$ at $\gamma_{BM} \ge 1$.

Fig.~\ref{Fig2}(b) presents $v_c(\gamma_{BM})$ dependence for different temperatures in double logarithmic scale. Inset shows these dependencies for $\gamma_{BM} \leq 1$. The characteristic voltage increases with the temperature decrease, in accordance with Fig.~\ref{Fig2}(a). For the commonly used temperature, $t \approx 0.5$, the characteristic voltage drops from its maximum value, $v_{c\max} \approx 0.46$, by 20\% at $\gamma_{BM} \approx 0.06$, and becomes halved at $\gamma_{BM} \approx 0.25$. Starting from $\gamma_{BM} \approx 1$, the characteristic voltage $v_c(\gamma_{BM})$ falls as $\gamma_{BM}^{-3/2}$ with $v_c < 0.1$. The decrease of the characteristic voltage with the suppression parameter increase is slower with the temperature decrease.

In the considered approximation, the characteristic voltage is nearly independent on the length of the bridge up to $L_{b}$ $\lesssim \xi _{n}$ for arbitrary $\gamma _{BM}$. The critical current, $I_{c}$, is the larger the smaller are $L_{b}$ and $\gamma _{BM}$. The current flowing through the junction ($I \approx I_c$) must not destroy the superconductivity in its electrodes. Below, we formulate the restriction on geometrical parameters of SN-N-NS structure coming from this requirement.

\section{Limitation on the current in superconducting electrodes  }

In accordance with experimental data obtained in MIT LL \cite{TolpygoASC}, a Nb strip with thickness, $d_{s}=200$~nm, and width, $W=250$~nm, possesses the critical current density, $J_{cst}\approx 4.5 \times 10^7$~A/cm$^2$ so that the critical current is $I_{cst} \approx 22.5$~mA at $T=4.2$ K. With an increase of the width, the critical current grows proportional to $\sqrt{W},$ while with $W$ decrease below 250~nm, $I_{cst}$ falls linearly with $W$,
\begin{equation}
I_{cst} = J_{cst}(W-W_{0})d_s\label{denJdep}
\end{equation}
where $W_{0}\approx 50$ nm is a doubled thickness of the contaminated surface layer.

The linear dependence of the critical current on the width means that in the interesting for us range of the thickness, $W\lesssim 250~$~nm, the supercurrent is distributed uniformly over the film width. This fact allows us to seek the solution of the Usadel equations in the depth of the electrodes $
(\left\vert x\right\vert \lesssim L)$ in the form $\Phi _{s,n}(x,y)=\Phi
_{s,n}(y)\exp \left\{ ikx\right\} $ independent on coordinate $z$ (see Fig.~\ref{Fig1}). Here $k$ is an independent on $\omega $ and $x$ constant, which is proportional to superfluid velocity. By assuming further that the supercurrent flowing through the SN electrodes is significantly less than the depairing current, we find functions $\Phi _{s,n}(x)$ from the solution of the proximity effect problem between superconducting and thin normal films,
\begin{equation*}
\Phi _{s}(x)=\Delta ,~\Phi _{n}(x)=\frac{G_{s}\Delta }{G_{s}+\omega \gamma
_{BM}},~G_{s}=\frac{\omega }{\sqrt{\omega ^{2}+\Delta ^{2}}}.
\end{equation*}
For supercurrent densities, $J_{S}$ and $J_{N}$, in the S and N films we get
\begin{equation}
\frac{e\rho _{s}J_{S}}{2\pi k_B T_{c}}=kS_{1},~S_{1}=t\sum_{\omega =0}^{\infty }%
\frac{\Delta ^{2}}{\omega ^{2}+\Delta ^{2}},~  \label{denJinS}
\end{equation}%
\begin{equation}
\frac{e\rho _{n}J_{N}}{2\pi k_B T_{c}}=kS_{2},~S_{2}=t\sum_{\omega =0}^{\infty }%
\frac{\Delta ^{2}}{\omega ^{2}(1+\omega \gamma _{BM}G_{s}^{-1})^{2}+\Delta
^{2}},  \label{denJinN}
\end{equation}%
where parameter $k$ in (\ref{denJinS}), (\ref{denJinN}),
\begin{equation}
k=\frac{e\rho _{s} I_{SN}}{2\pi T_{c}d_{s}WS_1 \left( 1+q\right) }, \label{k}
\end{equation}%
is determined by the magnitude of the full current, $I_{SN}=W\left(
J_{S}d_{s}+J_{N}d_{n}\right) $, flowing through the SN electrodes, while $q$ in (\ref{k}) is the ratio of the currents flowing through the N and S films, 
\begin{equation}
q= \frac{J_N d_n}{J_S d_s} =\frac{d_{n}\rho
_{s}}{d_{s}\rho _{n}}~\frac{S_{2}}{S_{1}},  \label{q}
\end{equation}
so that $1 + q = I_{SN}/J_S d_s W$.

At $T\approx 0.5T_{c}$, the Matsubara frequencies $\omega \ $are larger than $%
\Delta .$ We can neglect $\Delta $ in comparison with $\omega $ when
estimating the magnitudes of sums $S_{1,2}$ in (\ref{q}) and get, 
\begin{equation*}
~q=\frac{d_{n}\rho _{s}}{d_{s}\rho _{n}(1+\gamma _{BM})^{2}}.~
\end{equation*}


The current flowing through the S film must be small in comparison with S-film critical current,
\begin{equation}
\frac{I_c}{I_{cst}}\frac{J_S d_s W}{I_{SN}} = \frac{I_c}{(1+q)I_{cst}} \lesssim \beta,  \label{demand}
\end{equation}
where $\beta \ll 1$ is the desired smallness coefficient. The magnitude of $I_{c}$ in (\ref{demand}) can be obtained from the data presented in Fig.~\ref{Fig2}, $I_c = v_c 2 \pi k_B T_c/e R_n$. Since $I_c \sim R_n^{-1} \sim d_n$ and $I_{cst} \sim d_s$, the restriction (\ref{demand}) transforms into
\begin{equation}
\frac{\mathcal{C}}{1+q}\frac{d_n}{d_s} \lesssim \beta,
\label{demandLarge}
\end{equation}
where the coefficient $\mathcal{C}$ for niobium is
\begin{equation}
\mathcal{C} = \frac{v_c \times 5~\text{mV} }{J_{cst} \rho_n (L_b + 2\xi_n \sqrt{\gamma_{BM}})}.
\label{Ccoef}
\end{equation}
We should note that for $\rho_n \approx \rho_s$ the ratio of the currents flowing through the N and S films is small, $q \ll 1$, and so $1 + q \approx 1$.

\section{Heat balance  }

\begin{figure}[b]
	\includegraphics[width=1\columnwidth,keepaspectratio]{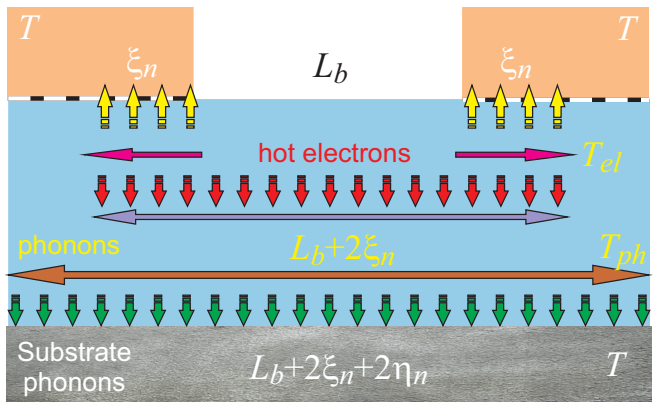}
	\caption{Sketch of an SN-N-NS Josephson junction with a variable thickness bridge geometry. The horizontal red, purple, and brown arrows indicate the areas where heat is transferred to the S electrodes, phonon subsystems in the normal film and substrate, respectively. The vertical yellow and red arrows schematically indicate the process of heat transfer from hot electrons to the electrodes and the phonon subsystem of a normal film. Vertical green arrows indicate the process of heat transfer into the substrate.}
	\label{Fig3}
\end{figure}

An important requirement to the design of circuits based on the SN-N-NS junction is that the normal component, $I_n$, of the total current must not lead to overheating of the SN-N-NS structure. Here we calculate the power dissipated in the junction during a clock period of a digital circuit and the power that can be absorbed by the junction environment. Then we evaluate the heating of niobium-copper SN-N-NS structure.

\subsection{Dissipated and absorbed power}

The heating effect arises due to Joule dissipation. The energy dissipated during the junction switching, $t_{sw} \approx \Phi_0/I_cR_n$, can be estimated as
\begin{equation}
E=\int_{0}^{t_{sw}}I_nVd \tau,  \label{power2}
\end{equation}
where $V=(\hbar /2e)\partial \varphi /\partial \tau$ is a voltage drop across the junction and $I_n = V/R_n$. By assuming that the phase increase rate is $\partial \varphi /\partial \tau \approx 2\pi/t_{sw}$ and that the junction is switching once in a clock period, $T_{clk}$, we obtain the dissipated energy and average dissipated power,
\begin{equation}
E \approx I_{c}\Phi _{0},~P=E/T_{clk}.  \label{disEn2}
\end{equation}

The energy $E$ is transferred to normal electrons located in the region of concentration of the electric field which includes the bridge between SN electrodes. The hot electrons may diffuse from the heated region on the characteristic time scale $\tau _{esc}\approx L_{b}^{2}/D_{n}.$ For a typical value of Fermi velocity, $v_{F}\approx 10^{6}$~m/s, electronic mean free path, $\ell \approx d_{n}\approx 10$~nm, and the bridge length, $L_b \approx \xi_n \approx 40$~nm, a rough estimation of $\tau _{esc}$ gives the value  $\sim 3\times 10^{-13}$~s. This is an order larger than the scattering time, $\tau_{sc} =\ell /v_{F}\approx 10^{-14}$~s.

Such a short value of $\tau _{esc}$ means that for a time equal to a clock period lying in picosecond time scale, hot electrons diffuse along the N film into the SN electrodes over a distance of the order of the current leakage length, $\xi_{n} \sqrt{\gamma _{BM}}$. This process is shown schematically by the red arrows in Fig.~\ref{Fig3}. 

Within the standard model of electron energy relaxation, the electron-electron collision rate is assumed to be large so that the electrons maintain an equilibrium distribution characterized by the temperature, $T_{loc}$, which is different from the lattice temperature $T$. The hot electrons may relax the excess energy making use of the two channels schematically shown in Fig.~\ref{Fig3} by yellow and blue arrows. 

The first channel for relaxation is the diffusion of hot electrons into superconducting electrodes. The heat conductance of the SN interface with small specific boundary resistance, $R_{B}$, is given by the expression \cite{Averin1996} 
\begin{equation}
P_{sn}=\frac{2\Delta }{e^{2}R_{B}}\mathit{A}\sqrt{\frac{2\pi \Delta }{%
k_{B}T_{loc}}}\exp \left( -\frac{\Delta }{k_{B}T_{loc}}\right)
k_{B}\delta T,  \label{kappaSN}
\end{equation}where  $\delta T=T_{loc}-T,$  $\mathit{A}$ is the effective area through which the heat flow takes place. For a finite transparency of SN interface $%
(\gamma _{BM}\approx 1)$, the area  $\mathit{A}\approx 2\xi _{n}W.$

The second channel is the transfer of excess energy to phonons. In dirty metals, the rate of this process , $P_{e-ph},$ is given by \cite{Belzig} 

\begin{equation}
P_{e-ph}=K_{\ast }\left( \frac{3\zeta (5)}{3\pi }\frac{c_{t}^{4}}{c_{l}^{4}}%
\frac{T_{loc}^{5}}{T_{\ast }^{5}}+\frac{4\pi ^{2}}{45}\frac{T_{loc}^{4}}{%
T_{\ast }^{4}}\right) ,~T_{loc}>T_{\ast },  \label{e-phpower}
\end{equation}%
\begin{equation}
K_{\ast }=\frac{N_{0}\tau_{sc} k_{B}T_{\ast }}{d_{fn}\ell ^{5}}\left( \frac{%
p_{F}^{2}}{m_e}\right) ^{2},~T_{\ast }=\frac{\hbar c_{t}}{k_{B}\ell }
\label{K}
\end{equation}%
where $N_{0}$ is the density of states at Fermy level, $c_{l},$ $c_{t},$ are longitudinal and transverse speeds of sound, $d_{fn}$ is the mass density, $p_{F}$ is Fermi momentum, $m_e$ is electron mass, $\zeta (z)$ is the Riemann zeta function. Expression (\ref{e-phpower}) can be rewritten via the thermal conductance per unit
volume $K(T)=dP_{e-ph}/dT_{loc}$ 
\begin{equation}
P_{e-ph}=\frac{K_{\ast }}{T_{\ast }}\left( \frac{15\zeta (5)}{3\pi }\frac{%
c_{t}^{4}}{c_{l}^{4}}\frac{T_{loc}^{4}}{T_{\ast }^{4}}+\frac{16\pi ^{2}}{45}%
\frac{T_{loc}^{3}}{T_{\ast }^{3}}\right) \delta T_{e-ph},  \label{e-phpower1}
\end{equation}where $\delta T_{e-ph} = $ $T_{loc}-T_{ph}$, $T_{ph},$ is the local temperature of phonon subsystem in the heated segment. 

The process of energy transformation from hot electrons to phonons occurs approximately in a volume $\left( L_{b}+2\xi _{n}\right) d_{n}W$ marked by a violet arrow in Fig.~\ref{Fig3}.  

Further phonon heat propagation along the N film can be described similar to the heat spread  in an infinite rod from the inner heated segment, $-\xi _{n}-L_{b}/2\leq x\leq \xi _{n}+L_{b}/2
$, which has a temperature $T_{loc}$ exceeding the temperature $T$ of its ends in the initial moment,
\begin{equation}
\begin{array}{c}
\delta T_{e-ph}(x,\tau)=\frac{\delta T_{e-ph}(x,0)}{2}\left[ \mathbf{erf}\left( 
\frac{x+\xi _{n}+L_{b}/2}{2\sqrt{\mu _{n}^{2}\tau}}\right) -\right.  \\ 
-\left. \mathbf{erf}\left( \frac{x-\xi _{n}-L_{b}/2}{2\sqrt{\mu _{n}^{2}\tau}}%
\right) \right] ,%
\end{array}
\label{deltaT1}
\end{equation}
where $\mathbf{erf}(z)$ is the error function, $\mu _{n}^{2}=\lambda
_{n}/C_{vn}d_{fn},$ $\lambda _{n}$ and $C_{vn}$ are thermal conductivity and thermal capacity of the N metal.

During a clock period, $\delta T_{e-ph}$ decreases from its initial value down to
\begin{equation}
\delta T_{e-ph}=\delta T_{e-ph}(x,0)\mathbf{erf}\left( \frac{2\xi
_{n}+L_{b}}{4\sqrt{\mu _{n}^{2}T_{clk}}}\right).  \label{deltaT2}
\end{equation}
The product $\eta _{n}=\mu _{n}\sqrt{T_{clk}}$ determines the characteristic scale of the heat propagation.

The excess phonon temperature may further relax to the substrate (see green arrows in Fig.~\ref{Fig3}). The power transfer from the phonons in the normal metal to the substrate is given by the Kapitza coupling \cite{Swartz}
\begin{equation}
P_{K}(T_{n};T)=K_{k}A_{ph}\left( T_{ph}^{4}(x,t)-T^{4}\right) \approx
4K_{k}A_{ph}T^{3}\delta T_{ph-s},  \label{kapitsa}
\end{equation}
where $\delta T_{ph-s}=T_{ph}-T,$  $K_{k}$ depends on the materials, and $A_{ph}\approx \left( L_{b}+2\xi _{n}+2\eta _{n}\right) W$ is the effective interface area of the heat transfer marked by brown arrow in Fig.~\ref{Fig3}.

\subsection{Heating of SN-N-NS structures based on Nb/Cu material combination}

The SN-N-NS  variable thickness bridge proposed in this work can be fabricated using commonly used materials like Nb, MoRe, V as a superconductor and Cu, Au, Al as a normal metal. Below, we estimate the parameters of NbCu-Cu-CuNb structure in accordance with the restriction (\ref{demandLarge}) and then calculate its heating assuming the clock period, $T_{clk} = 40$~ps (clock frequency, $f_{clk} = 25$~GHz). 

Substitution of the typical values \cite{VNbCu} of normal resistivity $\rho _{n}=3.7$~$\mu \Omega $ cm, decay length $\xi _{n}=37$ nm for Cu and $\rho _{s}=8$~$\mu \Omega $ cm for Nb into the obtained restriction on the critical current (\ref{demandLarge}) shows that the inequality can be fulfilled already for $\gamma_{BM} \geq 0.2$ at $T \approx 0.5 T_c$ with $\beta = 0.05$ if $L_b = \xi_n$, $d_n = 10$~nm, and $d_s \gtrsim 200$~nm. Here $\gamma_B = \gamma_{BM} \xi_n/d_n\approx 1$ is naturally implemented at the interface of the considered materials.

For $\gamma_{BM} = 0.5$ we obtain $v_c \approx 0.14$ (see Fig.~\ref{Fig2}) so that $I_cR_n = 0.7$~mV, $I_c \approx 0.53$~mA and $R_n \approx 1.3$~$\Omega$. The dissipated energy during a clock period is $E = I_c \Phi_0 \approx 1.1$~aJ and therefore the dissipated power is
\begin{equation}
P=E/T_{clk}\approx 2.7\times 10^{-8}~~\text{W}. 
\label{powerEl}
\end{equation}
The specific boundary resistance is $R_{B} =
\gamma_{BM}\rho _{n}\xi^2 _{n}/d_n\approx 2.5\times 10^{-15}~\Omega~$m$^{2}$. By putting further $T_{loc}\approx T = 4.2$ K and $\Delta \approx 1.69k_{B}T_{c}$ in (\ref{kappaSN}), we get the power flowing to the S films across SN interfaces,
\begin{equation}
P_{sn}=K_{sn} \delta T, 
\label{Psn}
\end{equation}where $K_{sn}\approx 2\times 10^{-7}$ W/K.

By substituting the typical Cu parameters, $v_{F}=1.57\times 10^{6}$~m/s, $
d_{fn}=8900$ kg/m$^{3},$ $c_{t}=4.8$ km/s, $c_{l}=2.3$ km/s, $T_{\ast }=1.8$~K, $\ell =10$ nm, $m_e = 9.1\times 10^{-31}$ kg, $E_{F}=1.13\times 10^{-18}$ J  into (\ref{e-phpower1}) and taking into account that the energy exchange between hot electrons and phonons occurs inside the volume of the order of $(L_b+2 \xi_n)d_nW$, we get the power transfer to phonons,
\begin{equation}
P_{e-ph}=K_{e-ph}\delta T_{e-ph},  
\label{Peph1}
\end{equation}
where $K_{e-ph}\approx 1.9\cdot 10^{-9}$ W/K.

Further power transfer between the normal metal and the substrate phonons is given by the Kapitza coupling (\ref{kapitsa}). The coefficient $K_{k}$ at Cu/Si interface approximately equals \cite{Ketchen1985,Clarke1989,Martinis1993} to $100$ W m$^{-2}$K$^{-4}$. By taking \cite{Novitskiy,Phillips,Pobell} the thermal conductivity, $\lambda _{n}\approx 500$ W/m~K, and thermal capacity, $C_{vn}\approx 0.3$ J/kg, for Cu, we get $\eta_n\approx 70$~nm and (\ref{kapitsa}) reads
\begin{equation}
P_{k}=K_{kk}\delta T_{ph-s},  
\label{Peph11}
\end{equation}
where $K_{kk}\approx1.5~10^{-9}$ W/K. 

In accordance with the heat balance described in the previous section, the power dissipated in the junction can be absorbed by the S-electrodes and the phonons of the N-film with subsequent partial power transfer into the substrate. Since $K_{kk} \approx K_{e-ph} \ll K_{sn}$,
one can estimate the heating of the structure as follows,

\begin{equation}
\delta T=\frac{P}{K_{sn}}\approx 0.13~\text{K.}  \label{deltaTel-bath}
\end{equation}%
Such temperature increase can lead to a slight decrease in the value of the critical current by the amount 
\begin{equation}
\left|\frac{\delta I_{c}}{I_{c}}\right|=\left|\frac{\partial v_c}{v_c \partial t}\right|
\frac{\delta T}{T_c}\approx 3.5\%,  \label{estimate}
\end{equation}see Fig.~\ref{Fig2}(a). The obtained deviation of the critical current seems acceptable in the view of standard optimization of the critical currents of Josephson junctions in digital circuits within the margins $\pm 20 \%$ and standard technological critical current spread, $\delta I_{ctec} \approx 3\%$.

The estimations (\ref{deltaTel-bath}), (\ref{estimate}) show that in the temperature range of interest, $T\approx 0.5T_{c}$, hot electrons effectively diffuse from the weak-link region into massive superconducting films, thereby eliminating the effect of the nonequilibrium state of the electronic subsystem on the mode of operation of SN-N-NS junction.  It is necessary to mention that the lowering of the operation temperature leads to exponential suppression of the heat transfer channel from hot electrons to S electrodes. This can lead to the noticeable difference between electron and phonon temperatures, especially in mK temperature range.

\section{Discussion}

In summary, we argue that SN-N-NS junctions with variable thickness bridge geometry are promising for miniaturization. The presented theoretical analysis shows that the junction $I_c R_n$ value reaches sub-millivolt level even when the rigid boundary conditions are not fulfilled at the SN interfaces. According to the obtained analytical expressions, the junction current-phase relation is close to a sinusoidal one at the operation temperature, $T/T_c \approx 0.5$. This allows the use of standard computer-aided design tools for superconducting digital circuit simulations. 

The suppression parameter value, $\gamma_{BM} \approx 1$, can be taken as the upper threshold for the fabrication of Josephson junctions intended for operation at liquid helium temperature, $T \approx 4.2$~K. The corresponding normalized characteristic voltage, $v_c \approx 0.08$, turns into $I_cR_n \approx 0.4$~mV and characteristic frequency, $I_cR_n/\Phi_0 \approx 200$~GHz. Since the clock frequency of complex circuits is usually a fraction of the characteristic one, the chosen value of $\gamma_{BM} \lesssim 1$ provides the possibility of operation with frequencies up to several tens of GHz.

Based on available experimental data \cite{TolpygoASC} we argue that the optimal width of SN-N-NS structure lays nearby $250$~nm. For a pair of materials like Nb and Cu, for $W=250$~nm, $d_n=10$~nm and $L_b=40$~nm, the characteristic voltage, $I_cR_n \approx 0.7$~mV, is achieved for $I_c\approx 0.53$~mA and $R_n\approx1.3~\Omega$.

An increase of $W$ leads to undesirable inhomogeneity of the supercurrent distribution in the S electrodes. Its decrease results in the fast suppression of the S-film critical current. The use of additional technological steps aimed at the decrease of SN interface transparency may be required to decrease the junction critical current density in this case.

One of possible ways of controllable reduction of the SN interface transparency is the technique that has proved itself in the fabrication of Nb/Al-AlO$_{x}$/Nb tunnel junctions \cite{Tolp,TolpSSJ,TolpSSJ1,MITech}. Here a Nb/Al sandwich can be used as a normal metal. A thin nonsuperconducting Nb film plays the role of a buffer layer. The subsequent growth of Al on such a buffer can contribute to the formation of an atomically smooth free surface of Al which provides predominantly mirror reflection of conducting electrons. The required Nb/Al boundary transparency, $\gamma_{B}$, can be achieved either naturally \cite{Zehnder1999,Brammertz2001} or by using the aluminum oxidation regime previously used in the fabrication of high-$J_{c}$ tunnel junctions \cite{TBGK2003}.  Contrary to double barrier devices, in SN-N-NS junctions the two SN interfaces are oxidised under the same conditions thus keeping the symmetry of the structure. 

The decrease of the transparency of the SN interfaces corresponds to a decrease in the characteristic voltage in proportion to $\gamma_{BM}^{-3/2}$ and in the critical current in proportion to $\gamma_{BM}^{-2}$. Ultimately, the width limitations is determined by the quality of the edges of the superconducting electrodes. In the presence of ``dead'' edges $\sim 25$~nm wide, the width can hardly be less than $150$~nm.

An effective area of SN-N-NS junction, $a_{JJ}$, can be determined according to the length within which the dominating heat transfer from hot electrons to superconducting films takes place, $L_{ht} = L_b + 2\xi_n$. Thus $a_{JJ} = 3\xi_n W$  for $L_b=\xi_n$ so that with $W=150$~nm we obtain $a_{JJ} \approx 18000$~nm$^2$. This is close to the area of semiconductor transistor, $a_t$, fabricated in the frame of a 40~nm technological process, under the assumption that $a_t \approx 50\lambda^2$, where $\lambda$ is the minimum feature size \cite{SemTrans}.  

\begin{acknowledgments}
The authors are grateful to D.V. Averin for fruitful discussions. Theoretical analysis of the considered SN-N-NS junction (Sections I - V) was supported by grant No. 20-12-00130 of the Russian Science Foundation. The study of the heat balance (Section VI) was performed according to the Development program of the Interdisciplinary Scientific and Educational School of Lomonosov Moscow State University "Photonic and Quantum technologies. Digital medicine",  and  the plan of scientific research  of the SINP MSU .
\end{acknowledgments}

\bibliography{NanoB2}

\end{document}



\title{Supplementary material to the article ``Miniaturization of Josephson junction for digital superconducting circuits''}

\author{I.~I.~Soloviev}
\email{isol@phys.msu.ru}
\affiliation{Lomonosov Moscow State
	University Skobeltsyn Institute of Nuclear Physics, 119991 Moscow,
	Russia} 
\affiliation{Dukhov All-Russia Research Institute of Automatics, Moscow 101000, Russia}
\affiliation{Physics Department of Moscow State University, 119991 Moscow, Russia}

\author{S.~V.~Bakurskiy}
\affiliation{Lomonosov Moscow State
	University Skobeltsyn Institute of Nuclear Physics, 119991 Moscow,
	Russia} 
\affiliation{Dukhov All-Russia Research Institute of Automatics, Moscow 101000, Russia}
\affiliation{Moscow Institute of Physics and Technology, State
	University, 141700 Dolgoprudniy, Moscow region, Russia}

\author{V.~I.~Ruzhickiy}
\affiliation{Lomonosov Moscow State
	University Skobeltsyn Institute of Nuclear Physics, 119991 Moscow,
	Russia}
\affiliation{Dukhov All-Russia Research Institute of Automatics, Moscow 101000, Russia}
\affiliation{Physics Department of Moscow State
	University, 119991 Moscow, Russia}

\author{N.~V.~Klenov}
\affiliation{Lomonosov Moscow State
	University Skobeltsyn Institute of Nuclear Physics, 119991 Moscow,
	Russia} 
\affiliation{Dukhov All-Russia Research Institute of Automatics, Moscow 101000, Russia}
\affiliation{Physics Department of Moscow State
	University, 119991 Moscow,
	Russia}

\author{M.~Yu.~Kupriyanov}
\affiliation{Lomonosov Moscow State
	University Skobeltsyn Institute of Nuclear Physics, 119991 Moscow,
	Russia} 

\author{A.~A.~Golubov}
\affiliation{Moscow Institute of Physics and Technology, State
	University, 141700 Dolgoprudniy, Moscow region, Russia}
\affiliation{Faculty of Science and Technology and MESA+ Institute of Nanotechnology, 7500 AE Enschede, The Netherlands}

\author{O.~V.~Skryabina}
\affiliation{Lomonosov Moscow State
	University Skobeltsyn Institute of Nuclear Physics, 119991 Moscow,
	Russia}
\affiliation{Moscow Institute of Physics and Technology, State
	University, 141700 Dolgoprudniy, Moscow region, Russia}

\author{V.~S.~Stolyarov}
\affiliation{Moscow Institute of Physics and Technology, State
	University, 141700 Dolgoprudniy, Moscow region, Russia}
\affiliation{Dukhov All-Russia Research Institute of Automatics, Moscow 101000, Russia}

\date{\today}

\maketitle

In this supplementary material we present the details of a SN-N-NS junction superconducting current calculation.

\section{Model of a SN-N-NS structure}

\begin{figure}[b]
\includegraphics[width=0.6\columnwidth,keepaspectratio]{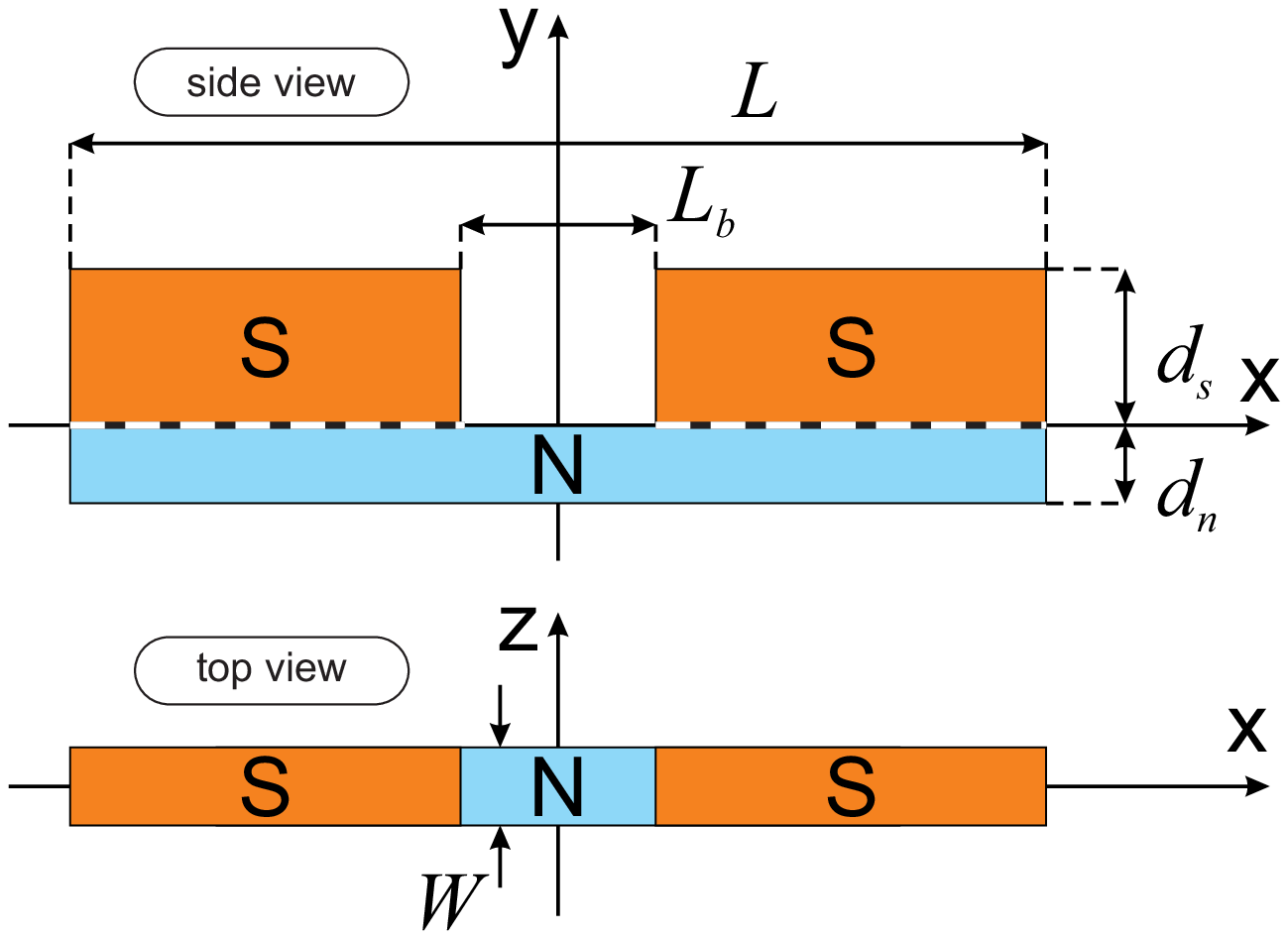}  
\caption{Sketch of a SN-N-NS Josephson junction with a variable thickness bridge geometry.}
\label{Fig1}
\end{figure}

The model of SN-N-NS junction contains a normal metal film connecting two massive superconducting electrodes of the length, $L-L_b,$ located at the distance $\pm L_b/2$ from the centre of this film, see Fig.~\ref{Fig1}. The existence of the proximity effect between N and S materials leads to the induction of superconducting correlations in the N film and, therefore, the Josephson effect between the S-banks.


In the calculation of the critical current, we suppose that the condition of the dirty limit is fulfilled for all metals, the critical temperature of the N-material is equal to zero and its width, $W$, and thickness, $d_{n}$, is much smaller than Josephson penetration depth, $\lambda_{J}$, and decay length, $\xi_{n}=\sqrt{D_{n}/2\pi T_{c}}$, respectively.

The proximity effect in this system can be considered in the frame of Usadel equations \cite{Usadel}, which in the N-film have the form
\begin{equation}
\frac{\partial }{\partial x}\left( G_{n}^{2}\frac{\partial \Phi _{n}}{\partial x}\right) +\frac{\partial }{\partial y}\left(G_{n}^{2}\frac{\partial \Phi_{n}}{\partial y}\right) =\omega G_{n}\Phi_{n}.
\label{equTetaNF}
\end{equation}
Here  $G_{n}=\omega /\sqrt{\omega^{2}+\Phi_n\Phi_{n}^{\ast }}$, $\Phi_{n}$ are modified Usadel Green's functions normalized on $\pi T_{c}$ , $\omega =(2m+1)T/T_{c}$ are Matsubara frequencies  normalized on $\pi T_{c}$ ($m$ is integer), $x$ and $y$ are coordinates along and across the N-film normalized on $\xi_{n}$. We place the origin of coordinates on the SN interface in the middle of the N-film, as it is shown in Fig.~\ref{Fig1}.

The equations (\ref{equTetaNF}) should be supplemented by the boundary conditions. Due to the symmetry of the problem, we can solve them for the positive $x$ only, making use of the boundary conditions
\begin{equation}
\frac{\partial }{\partial x}\re\Phi_{n}=0,\quad \im\Phi_{n}=0
\label{BcCenter}
\end{equation}
in the center of the bridge, $x=0$. Here $\re\Phi_{n}$ and $\im\Phi_{n}$ are the real and imaginary parts of the functions $\Phi_{n}.$

At the free N-interfaces $(0<x<L/2,$ $y=-d_{n})$, $(0<x\leq L_b/2,$ $y=0)$ the boundary conditions are
\begin{equation}
\frac{\partial }{\partial y}\Phi_{n}=0,  \label{BCfree}
\end{equation}%
and at $x=L/2$, 
\begin{equation}
\frac{\partial }{\partial x}\Phi_{n}=0.  \label{BcInf}
\end{equation}%
At the SN interface $(L_b/2\leq x\leq L/2,~y=0)$ the functions $\Phi _{n}$ and the Usadel Green's functions in the S-electrode, $\Phi_{s},$ should be coupled
by the Kupriyanov-Lukichev boundary condition \cite{kuprlukichevA} 
\begin{equation}
\gamma_{B}G_{n}\frac{\partial }{\partial y}\Phi_{n}=G_{s}\left( \Phi_{s}-\Phi_{n}\right),  \label{BCgammaB}
\end{equation}
where $\gamma_{B}=R_{B}/\rho_{n}\xi_{n}$ is suppression parameter, $R_{B}$ is specific resistance of the SN interface, $G_{s}=\omega\sqrt{\omega^{2}+\Phi_{s}\Phi_{s}^{\ast}}$.

The conditions \cite{Golubov1983} 
\begin{equation}
\gamma_{M}\lesssim 0.3,\quad d_{n}\ll 1  \label{restr}
\end{equation}
permit us to neglect the suppression of superconductivity in the S-film due to the proximity effect even in the case of a fully transparent SN interface. We consider the Usadel Green's functions, $\Phi_{s}$, equal to their equilibrium values in a superconductor at a given temperature, 
\begin{equation}
\Phi_{s}=\Delta \exp \{i\varphi /2\},\quad G_{s}=\frac{\omega }{\sqrt{\omega^{2}+\Delta^{2}}},  \label{tetaSinS}
\end{equation}
correspondingly. Here $\Delta $ is the magnitude of superconducting order parameter normalized on $\pi T_{c}$, $\varphi $ is the superconducting phase difference across the junction.

The conditions (\ref{restr}) also allow considering functions $\Phi_{n}$ independent of the coordinate $y$, in the first approximation on the
parameter $d_{n}$. By integration of the equation (\ref{equTetaNF}) over the coordinate $y$, and using the boundary conditions (\ref{BCfree}), (\ref{BCgammaB}), we obtain the equations 
\begin{equation}
\frac{\gamma_{BM}}{G_{n}}\frac{\partial }{\partial x}\left( G_{n}^{2}\frac{\partial \Phi_{n}}{\partial x}\right) -\left( G_{s}+\gamma_{BM}\omega\right) \Phi_{n}=-G_{s}\Phi_{s},  \label{EqUnderSr}
\end{equation}
in the region $L_b/2\leq x\leq L/2$, where $\gamma_{BM}=\gamma_{B}d_{n}$.

In the area $0\leq x\leq L_b/2$ we have the equations 
\begin{equation}
\frac{\partial }{\partial x}\left( G_{n}^{2}\frac{\partial \Phi_{n}}{\partial x}\right) =\omega G_{n}\Phi_{n}.  \label{EqOutS}
\end{equation}
The solution of equations (\ref{EqUnderSr}), (\ref{EqOutS}) is simplified in the limit of a small gap between the superconducting electrodes, $L_b\ll 1$.

The obtained functions $\Phi _{n}$ allows to calculate the supercurrent $J_{s}$ across the junction
\begin{equation}
\frac{eRJ_{s}}{2\pi T_{c}}=t\sum_{\omega \geq 0}\frac{G_{n}^{2}}{\omega ^{2}}%
Re\Phi _{n}\frac{\partial }{\partial x}Im\Phi _{n},\quad R=\frac{\rho
_{n}\xi _{n}}{d_{n}W},  \label{currentDef}
\end{equation}%
where $t = T/T_c$ is the normalized temperature.

\section{Solution of the problem}

\subsection{The bridge between the S-electrodes}

In the first approximation on $L_b \ll 1$, we may neglect the right-hand side of the equations (\ref{EqOutS}), 
\begin{equation}
\frac{\partial }{\partial x}\left( G_{n}^{2}\frac{\partial \Phi_{n}}{\partial x}\right) =0,  \label{outS1}
\end{equation}
in the area $0\leq x\leq L_b/2$.

The functions $\re\Phi_{n}$ are independent of the coordinate $x$, 
\begin{equation}
\re\Phi_{n}=\re\Phi_{n}(L_b/2),  \label{outS4}
\end{equation}
as it follows from the boundary conditions (\ref{BcCenter}), the condition of continuity of the functions $\Phi _{n}$ at $x=L_b/2$, and the equations (\ref{outS1}).

For the imaginary part of $\Phi_{n}$ we obtain 
\begin{equation}
G_{n}^{2}\frac{\partial }{\partial x}\im\Phi_{n}=\frac{2}{L_b}\frac{\omega^{2}}{\sqrt{\omega^{2}+(\re\Phi_{n})^{2}}}C,  \label{outS6}
\end{equation}
\begin{equation*}
G_{n}^{2}=\frac{\omega^{2}}{\omega^{2}+(\re\Phi_{n})^{2}+\im\Phi_{n}^{2}},
\end{equation*}
where $C$ are integration constants. By taking into account that $\re\Phi_{n}$ are constants (\ref{outS4}), we may find the solution of (\ref{outS6}) in the form 
\begin{equation}
\im\Phi_{n}=\sqrt{\omega^{2}+(\re\Phi_{n})^{2}}\tan \theta. \label{outS7}
\end{equation}
Substitution of (\ref{outS7}) into (\ref{outS6}) leads to
\begin{equation}
\frac{\partial \theta }{\partial x}=\frac{2}{L_b}C.  \label{outS8}
\end{equation}
This results in a linear dependence of the functions $\theta$ on the coordinate $x$, 
\begin{equation}
\theta =\frac{2}{L_b}Cx,  \label{outS9}
\end{equation}
so the equations for the imaginary part of $\Phi_{n}$ are 
\begin{equation}
\im\Phi_{n}=\sqrt{\omega^{2}+(\re\Phi_{n})^{2}}\tan \frac{2C}{L_b}x.
\label{out10}
\end{equation}

\subsection{The regions under the S-electrodes}

According to the equations (\ref{EqUnderSr}), for the real part of $\Phi_{n}$ under the S-electrodes we have 
\begin{equation}
\xi_{ef}^{2}\frac{\partial }{\partial x}\left( G_{n}^{2}\frac{\partial }{\partial x}\re\Phi_{n}\right) - \re\Phi_{n}=-\delta \cos \frac{\varphi }{2},  \label{underS1}
\end{equation}
where 
\begin{equation*}
\delta =\frac{G_{s}\Delta }{\left( G_{s}+\gamma_{BM}\omega \right) },~~~\xi_{ef}=\sqrt{\frac{\gamma_{BM}}{G_{n}\left( G_{s}+\gamma_{BM}\omega \right)}}.
\end{equation*}
The equations (\ref{underS1}) should be solved with the boundary conditions, 
\begin{equation}
\frac{\partial }{\partial x}\re\Phi_{n}=0,
\end{equation}
at $x=L/2$ and $x=L_b/2.$

The solutions of this boundary problems are obviously constants independent of $x$, 
\begin{equation}
\re\Phi_{n}=\delta \cos \frac{\varphi }{2}.  \label{ReFn}
\end{equation}

The equations for the imaginary parts of $\Phi_{n}$ can be written as follows, 
\begin{equation}
\xi_{ef}^{2}\frac{\partial }{\partial x}\left( G_{n}^{2}\frac{\partial }{\partial x}\im\Phi_{n}\right) -\im\Phi_{n}=-\delta \sin \frac{\varphi }{2},
\label{underS2}
\end{equation}
where 
\begin{equation}
G_{n}=\frac{\omega }{\sqrt{\omega^{2}+\left( \re\Phi_{n}\right)^{2}+\left( \im\Phi_{n}\right) ^{2}}}.  \label{underS3}
\end{equation}
By taking into account that $\re\Phi_{n}$ are constants (\ref{ReFn}), we again may find the solution of equation (\ref{underS3}) in the form (\ref{outS7}). Its substitution into (\ref{underS3}) leads to 
\begin{eqnarray}
G_{n} &=&\frac{\omega \cos \theta }{\sqrt{\omega^{2}+\left(\re\Phi_{n}\right)^{2}}}.
\end{eqnarray}
This results in the following equation for the $\theta$ functions, 
\begin{equation}
\frac{\gamma_{BM}\sqrt{\omega^{2}+\Delta^{2}}}{\Omega_{1}}\frac{\partial^{2}\theta }{\partial x^{2}}-\sin \theta =-\frac{\Delta \sin \frac{\varphi }{2}}{\Omega_{1}}\cos \theta,  \label{underS6}
\end{equation}
where 
\begin{equation}
\Omega_{1}=\sqrt{\Omega^{2}+\Delta^{2}\cos^{2}\frac{\varphi}{2}},~~~\Omega = \omega \left( 1+\gamma_{BM}\sqrt{\omega^{2}+\Delta^{2}}\right). \label{underS7}
\end{equation}

\subsection{General solution}

The first integral of the equation (\ref{underS6}) gives
\begin{equation}
\frac{\gamma_{BM}\sqrt{\omega^{2}+\Delta^{2}}}{2\Omega_{1}}\left( \frac{\partial \theta }{\partial x}\right)^{2}+\cos \theta -\cos \Theta = \frac{\Delta \left( \sin \Theta -\sin \theta \right) \sin \frac{\varphi }{2}}{\Omega_{1}},  \label{un1} \end{equation}
where $\Theta$ are the magnitudes of the functions $\theta (x)$ at $x=L/2.$

By matching this solution at $x=L_b/2$ with the one (\ref{outS9}) in the area $0 \leq x\leq L_b/2$, we obtain an equation with respect to constants $C$, 
\begin{equation}
\frac{2\eta^{2}\sqrt{\omega^{2}+\Delta^{2}}}{\Omega_{1}}C^{2}+\cos C-\cos \Theta = \frac{\Delta \left( \sin \Theta -\sin C\right) \sin \frac{\varphi }{2}}{\Omega _{1}}, \label{underS8}
\end{equation}
which, in turn, determines the product of the superconducting current, $J_{s}$, across the junction and its normal state resistance, $R_n$, 
\begin{equation}
\frac{eR_{n}J_{s}}{2\pi T_{c}}=t\sum_{\omega \geq 0}\frac{2\re\Phi_{n}C}{\sqrt{\omega^{2}+(\re\Phi_{n})^{2}}}\left( 1+2\eta \coth \frac{L-L_b}{2\sqrt{\gamma_{BM}}}\right).  \label{Js}
\end{equation}
Here $\eta =\sqrt{\gamma_{BM}}/L_b$, while 
\begin{equation}
R_{n}=R_{nb}+2R_{sn}
\end{equation}
is the sum \cite{Karminskaya2010} of the resistance of the bridge, $R_{nb}$, and the resistances of two SN interfaces, $R_{sn}$, 
\begin{equation*}
R_{nb}=\frac{\rho _{n}L_b}{d_{n}W},~~~R_{sn}=\frac{\rho _{n}\xi _{n}\sqrt{\gamma_{BM}}}{d_{n}W}\coth \frac{L-L_b}{2\xi_{n}\sqrt{\gamma_{BM}}}.
\end{equation*}

From the equations (\ref{underS6}) and (\ref{un1}) it follows that the characteristic scale, $\zeta$, of spatial variation of the functions $\theta$ along the coordinate $x$ is 
\begin{equation}
\zeta =\sqrt{\frac{\gamma_{BM}\sqrt{\omega^{2}+\Delta^{2}}}{2\sqrt{\omega^{2}\left( 1+\gamma_{BM}\sqrt{\omega^{2}+\Delta^{2}}\right)^{2}+\Delta^{2}\cos^{2}\frac{\varphi}{2}}}}.  \label{ScaleX}
\end{equation}
At $t\gtrsim 0.5$ the sum in (\ref{Js}) converges at $\omega >\Delta$. By neglecting $\Delta$ in comparison with $\omega$ in (\ref{ScaleX}) we obtain the following estimation for $\zeta$, 
\begin{equation}
\zeta =\sqrt{\frac{\gamma_{BM}}{2\left( 1+\gamma_{BM}\omega \right) }}.
\label{scaleTc}
\end{equation}
At small temperatures, the sum in (\ref{Js}) converges at $\omega \approx \Delta $. According to (\ref{ScaleX}), we obtain 
\begin{equation}
\zeta \approx \sqrt{\frac{\gamma_{BM}}{\sqrt{2}\sqrt{\left( 1+\gamma_{BM}\Delta \sqrt{2}\right)^{2}+\cos^{2}\frac{\varphi }{2}}}}.
\label{scale0}
\end{equation}
From (\ref{ScaleX}) - (\ref{scale0}) it follows that in the limit of small $\gamma_{BM}$ the characteristic scale is $\zeta $ $\approx $ $\sqrt{\gamma_{BM}},$ while for a large $\gamma_{BM}$ it is $\zeta \approx \Delta^{-1/2}$ $\approx 1$ at small temperatures and $\zeta \approx \omega^{-1/2}$ at $t\gtrsim 0.5$.

In a practically interesting case, the length of the superconducting electrodes is much larger than the characteristic scale, 
\begin{equation}
L-L_b\gg \zeta.  \label{largeLS}
\end{equation}
Here the magnitude of the functions $\theta (x)$ at $x=L/2$ can be easily found from (\ref{un1}) as 
\begin{equation}
\tan \Theta =\frac{\Delta \sin \frac{\varphi }{2}}{\sqrt{\Omega^{2}+\Delta^{2}\cos^{2}\frac{\varphi}{2}}},  \label{tanTeta}
\end{equation}
resulting in 
\begin{equation}
\sin \Theta =\frac{\Delta \sin \frac{\varphi }{2}}{\sqrt{\Omega^{2}+\Delta^{2}}},  \label{sinTeta}
\end{equation}
\begin{equation}
\cos \Theta =\frac{\sqrt{\Omega^{2}+\Delta^{2}\cos^{2}\frac{\varphi }{2}}}{\sqrt{\Omega^{2}+\Delta^{2}}}.  \label{cosTeta}
\end{equation}
The expression (\ref{Js}) here has the form 
\begin{equation}
\frac{eR_{n}J_{s}}{2\pi T_{c}}=t\sum_{\omega \geq 0}\frac{2\re\Phi_{n}\left( 1+2\eta \right) }{\sqrt{\omega^{2}+(\re\Phi_{n})^{2}}}C.
\label{Js1}
\end{equation}
Substitution of (\ref{sinTeta}) and (\ref{cosTeta}) into (\ref{underS8}) leads to the equations which are closed relatively the constants $C$. Their analytical solutions in the limits of small and large values of the suppression parameter $\gamma_{BM}$ are presented in the main text of the paper.

\bibliography{NanoB2}